\documentclass[submission,copyright,creativecommons]{eptcs}
\usepackage{amssymb, amsmath,amsthm}
\usepackage{xspace}
\usepackage{tikz}

\providecommand{\event}{ARCADE 2019}

\newtheorem{definition}{Definition}[section]
\theoremstyle{remark}
\newtheorem{example}{Example}[section]

\newcommand{\highlight}[2][green!80!black!50!yellow!40]{%
\setlength{\fboxsep}{0.1mm}\setlength{\fboxrule}{0mm}%
\colorbox{#1}{\:\ensuremath{#2\strut}\:}}
\newcommand{\BLUE}[1]{\highlight[blue!80!green!80!black!20]{#1}}
\newcommand{\GREEN}[1]{\highlight[green!80!blue!70!black!20]{#1}}
\newcommand{\RED}[1]{\highlight[red!80!yellow!80!black!20]{#1}}

\newcommand{\YELLOW}[1]{\highlight[red!30!yellow!80!black!30]{#1}}

\newcommand{\tool}[1]{\textsf{#1}}

\newcommand{\RR}{R}

\newcommand{\semantic}{structural}
\newcommand{\depgraph}{clause dependency graph\xspace}
\newcommand{\depgraphs}{clause dependency graphs\xspace}
\newcommand{\secref}[1]{Section~\ref{sec:#1}}
\newcommand{\m}[1]{\mathsf{#1}}
\newcommand\etal{et al.\xspace}

\title{Smarter Features, Simpler Learning?}
\author{
Sarah Winkler
\institute{University of Verona, Italy}
\email{sarahmaria.winkler@univr.it}
\and
Georg Moser
\institute{University of Innsbruck, Austria}
\email{georg.moser@uibk.ac.at}
}

\def\authorrunning{S. Winkler and G. Moser}
\def\titlerunning{Smarter Features, Simpler Learning?}

\begin{document}

\maketitle

\begin{abstract}
Earlier work on machine learning for automated reasoning mostly relied on
simple, syntactic features combined with sophisticated learning techniques.
Using ideas adopted in the software verification community,
we propose the investigation of more complex, {\semantic}
features to learn from. These may be exploited to either learn beneficial strategies
for tools, or build a portfolio solver that chooses the most suitable tool for 
a given problem.
We present some ideas for features of term rewrite systems and theorem
proving problems.
\end{abstract}

\section{Introduction}
\label{sect:introduction}

The idea to exploit machine learning in automated reasoning has been a natural one,
given the success of AI methods in areas like Jeopardy%
\footnote{\textsc{Watson}, IBM, %
\url{https://researcher.watson.ibm.com/researcher/view_group.php?id=2099}}
and Go.%
\footnote{\textsc{AlphaGo}, DeepMind, %
\url{https://deepmind.com/research/alphago}.}
Indeed, machine learning has already been investigated to improve
heuristics for different tasks in theorem proving.
Most prominently, these include learning parameters of the proof process~%
\cite{BHP14},
learning selection of the objects (like clauses) to process next~%
\cite{Sch01,JU17},
premise selection~\cite{JU18},
and learning proof search guidance in a more general sense,
such as which branch to follow in a tableau proof~\cite{KU15}.%

Work done in the area so far typically exploited features reflecting the
syntactic structure of the input; we mention some examples as representatives
of different approaches.
Used characteristics commonly include symbol counts, the number of clauses of
a certain type, clause depth, and clause weight. For instance, such properties
were considered by Bridge \etal~\cite{BHP14} to determine tool parameters both for the initial 
problem (\emph{static} features) and during the proof process (\emph{dynamic} features).
The strategy scheduling learner E-MaLeS by Kaliszyk \etal~\cite{KSU13} also relies
on features such as clause, literal, and term count, function symbol arity, and
problem class (e.g., Horn or unit problems).
Another approach used for clause selection in early work by Schulz~\cite{Sch01}
and Jakub\r{u}v and Urban~\cite{JU17,CJSU19} is to extract symbol strings
(sometimes called \emph{term walks}) from the tree representation of terms, and
count occurrences of a predefined set of strings to obtain a feature vector.
In a further experiment by Loos \etal~\cite{LISK17} the entire problem was fed
into a neural network.
In the context of premise selection for a given conjecture, terms in the
conjecture have often been compared to the terms in the premises, using the
well-known notion of term frequency-inverse document frequency, but also
by comparing prefixes of terms, see e.g. work by Jakub\r{u}v and Urban~\cite{JU18}.
\smallskip

These approaches tend to use elaborate machine learning models with rather
simple features. A contrasting approach following the paradigm of \emph{smart
features, simple learning} has been pursued by Demyanova \etal to 
design a portfolio solver for the software verification competition (SV-COMP)
\cite{DPVZ17}:
Based on metrics of programs related to loop types, variable use, and control
flow, the authors applied machine learning techniques to decide which solver
to use. The resulting portfolio solver would have won the software competition
in three consecutive years.
Since the applied machine learning model is comparatively simple (support vector
machines), the success of the method is attributed to the sophisticated
metrics.
Besides providing a powerful combined solver, the work thus also delivers a
taxonomy of problems, along with insights on which techniques and tools work
best on them.
\smallskip

We propose to experiment with a similar approach to the areas of term rewriting
and theorem proving.
Both in the termination%
\footnote{\url{http://termination-portal.org/wiki/Termination\_Competition}}
and the confluence competition\footnote{\url{http://project-coco.uibk.ac.at}}
for term rewrite systems
a variety of tools compete annually on thousands of benchmarks. Each tool has
its own strengths on problems of a certain kind, but it can typically also be
run with a vast variety of different strategies that accommodate better to
some problems than others.
We do not deny the relevance of syntactical characteristics: some tools and 
analysis techniques are obviously restricted to problems with certain syntactic
properties. For instance, some tools are restricted to string rewrite systems,
and the Knuth-Bendix order only applies to non-duplicating rewrite systems.
However, many tools work on a wide range of problems and offer a variety
of proof strategies. This holds for rewrite tools that attempt to establish
termination, complexity bounds, and confluence; but also for theorem
provers competing in the CADE Annual System Competition (CASC).%
\footnote{\url{http://www.tptp.org/CASC}}
It is a common experience in this field that with the ``right'' strategy, a
proof can be found within a few seconds or even fractions thereof while
other strategies diverge hopelessly; but a suitable strategy is hard to predict.
Many tools thus employ \emph{strategy scheduling}, where a variety of strategies
is run subsequently on a problem for a short time each, in the hope that one of
them might succeed~\cite{KSU13}.
\smallskip

In this extended abstract we thus propose to investigate meaningful {\semantic}
features that, possibly together with syntactic characteristics, allow machine
learning models to predict which proof strategy and/or tool works best for a
given problem.
We do not fix a particular learning task, as we think that such {\semantic}
properties could be of interest for different concrete applications.

\smallskip
In \secref{programs} we summarize the program metrics exploited
in~\cite{DPVZ17}.
We propose corresponding metrics for term rewrite systems in \secref{trs}.
Subsequently, we consider possible extensions to problems from theorem proving
in \secref{tp} and conclude in \secref{conclusion}.

\section{Program Metrics}
\label{sec:programs}

The metrics to classify programs proposed in~\cite{DPVZ17} are related to three
different aspects: variable roles, loop patterns, and control flow.

For the former, the authors defined 27 variable roles, including e.g., pointers,
loop bounds, array indices, counters, and variables of a particular type.
An occurrence count vector holding the number of variables of each role,
normalized by the total number of variables, is included in the features used
for learning.

Second, four loop patterns are defined based on an estimate whether the number
of iterations in a program execution can be bounded or not.
In particular, the first two patterns correspond to loops with
a constant number of iterations, and a certain type of FOR-loops where
termination is guaranteed.
Again, the number of loops of each kind gets counted, and an occurrence count
vector (normalized by the number of loops overall) is included in the feature
list.

Finally, control flow metrics include the number of basic blocks and the
maximal indegree of a basic block for intraprocedural control flow, as well
as the number of (recursive) call expressions and the involved parameters to
account for interprocedural control flow.

\section{Metrics for Rewrite Systems}
\label{sec:trs}

In this section we propose structural features for rewrite systems like
the ones collected in the Termination Problems Database \tool{TPDB}%
\footnote{\url{http://termination-portal.org/wiki/Termination\_Competition\#Termination\_Problems\_Data\_Base}} or the 
collection of confluence problems \tool{CoPs}.%
\footnote{\url{http://termcomp-devel.uibk.ac.at/cops}}
Some basic notions of term rewriting are assumed in the sequel, see e.g.~\cite{BN98}.

As mentioned above, there are a number of syntactic features of rewrite
systems which naturally restrict applicable tools and
strategies. These include the type of rewrite systems (string, conditional,
constrained, classes of higher-order systems). 
On the other hand, one may want to recognize variable duplication (e.g. for
the Knuth-Bendix order), linearity (e.g. for confluence criteria), or
left-linear right ground systems (whose first-order theory is known to be
decidable~\cite{DT90}).

That said, we next discuss structural features which might influence
applicability of certain tools and methods, as outlined in Section~%
\ref{sec:programs}.
While programs are generally more structured than rewrite systems, we
argue in the next paragraphs that several properties outlined in
\secref{programs} actually remain relevant.

\paragraph{Variable Roles.}
On the level of term rewrite systems, a program variable corresponds to an
\emph{argument position} of a function symbol.
We consider a function symbol $f$ of arity $n$ occurring in a TRS $\RR$, with argument 
position $1 \leq i \leq n$.
The following properties of the $i$th argument can be considered:
\begin{itemize}
\item it is a \emph{projection argument} if $\RR$ contains a
rule $f(t_1,\dots,t_n) \to t_i$,
\item it is \emph{decreasing} in presence of a rule
$f(t_1,\dots,C[t_i],\dots t_n) \to f(t_1,\dots,t_i,\dots t_n)$
where $C$ is a non-empty context
(or \emph{increasing} if the rule is reversed),
\item it is \emph{recursive} if some rule
$f(t_1,\dots, t_n) \to f(t_1,\dots,f(u_1,\dots, u_n),\dots t_n)$ features
a recursive call in the $i$th position,
\item it is a \emph{pattern matching argument} if $\RR$ has rules
$f(t_1,\dots,c_1,\dots, t_n) \to r_1$ and $f(t_1,\dots,c_2,\dots, t_n) \to r_1$ with different
constructors $c_1$ and $c_2$ occurring at position $i$, and
\item it is a \emph{duplication argument} if in a rule
$f(t_1,\dots, t_n) \to r$ term $t_i$ occurs at least twice in $r$.
\end{itemize}
Some of these properties resemble roles of program
variables: for instance, a decreasing or increasing argument position can be
seen as the equivalent of a counter variable in a TRS, and a projection argument
corresponds to a variable that is returned by a function.
In the case of integer transition systems, many more of the variable roles
from~\cite{DPVZ17} have straightforward equivalents, for instance counters,
argument positions involved in branching conditions, and arguments involved
in arithmetic operations .

Moreover, for particular types of rewrite systems, further roles become
important:
In higher-order systems, argument positions can be divided into first-
and higher-order ones. In the case of conditional or constrained TRSs,
positions having
\emph{extra variables} can be recognized (i.e., variables that occur in
the right-hand side of a rule but not in the respective left-hand side) .

\paragraph{Recursion Patterns.}
In particular for termination and complexity analysis of rewrite systems,
recursion constitutes the key difficulty.
The first two loop patterns considered in~\cite{DPVZ17} are strongly connected
to the idea of tiering and safe
recursion~\cite{Si88,BellantoniCook:1992,Leivant:1994}, where the essential
idea is that the arguments of a function are separated into normal and safe
arguments, and recursion is only allowed in safe arguments.
Such a problem classification is also commonly used in runtime complexity
analysis of term rewrite systems~\cite{AM13,AEM15}.
More generally, restricted loop structures have been considered in the context
of decidable classes of resource or termination analysis,
cf.~\cite{Ben-Amram:2011,BP:2016,CDZ:2017}.
Again, for the case of integer transition systems the loop patterns
of~\cite{DPVZ17} naturally carry over.

\paragraph{Control Flow.}
Just like call graphs depict calling relationships between functions in
imperative programs, a call graph can be considered for term rewrite systems,
and e.g. recursive calls and degrees of nodes can be counted to obtain numerical
features.

The dependency graph~\cite{AG00} (DG) is another common illustration of control
flow in TRSs. This graph is in general not computable, but estimations thereof 
are often used in termination tools.
Natural features of this graph are given by the number of nodes, edges,
and strongly connected components (SCCs).
Though the control flow is in general less obvious for TRSs than for
programs, some of the  program metrics mentioned above have 
quite direct correspondents: The number of recursive calls is reflected
by the number of paths in the DG between nodes with the same root
symbol. Mutually recursive functions are reflected by edges between 
nodes and SCCs of different root symbols. The indegree of nodes in the
DG could be considered related to the indegree of basic blocks.
%

We illustrate some of the proposed properties by means of an example.

\begin{example}
\label{exa:eratosthenes}
The following TRS $\RR$ models a functional program to enumerate prime
numbers:
\begin{xalignat*}{4}
&(1)&
\m{primes} &\to \m{sieve}(\m{from}(\m{s}(\m{s}(\m{0})))) &
&(5)&
\m{sieve}(\m{0}:y) &\to \m{sieve}(y) \\
&(2)&
\m{from}(n) &\to n:\m{from}(\m{s}(n)) &
&(6)&
\m{sieve}(\m{s}(n):y) &\to \m{s}(n):\m{sieve}(\m{filter}(n,y,n)) \\
&(3)&
\m{take}(\m{0},y) &\to \m{nil} &
&(7)&
\m{filter}(\m{0},x:y,m) &\to \m{0}:\m{filter}(m,y,m) \\
&(4)&
\m{take}(\m{s}(n),x:y) &\to x:\m{take}(n,y) &
&(5)&
\m{filter}(\m{s}(n),x:y,m) &\to x:\m{filter}(n,y,m)
\end{xalignat*}
For instance, the first arguments of $\m{take}$ and $\m{filter}$ are
both decreasing and pattern matching arguments,
while the argument of $\m{from}$ is increasing. Both can be
seen as analogues of counter variables. 
The last argument of $\m{filter}$ is a duplication argument.

The call graph of this program depicts calling
relationships between the different functions: an edge from function $f$ to
function $g$ indicates that there exists a rule with $f$ as the root symbol
of a left-hand side where $g$ occurs on the corresponding right-hand side.
\begin{center}
\begin{tikzpicture}[node distance=16mm]
\node (3) {$\m{take}$};
\node[right of=3] (2) {$\m{from}$};
\node[right of=2] (1) {$\m{primes}$};
\node[right of=1] (4) {$\m{sieve}$};
\node[right of=4] (5) {$\m{filter}$};
\draw[->] (1) -- (2);
\draw[->] (1) edge (4);
\draw[->] (2) edge[loop below] (2);
\draw[->] (3) edge[loop below] (3);
\draw[->] (4) -- (5);
\draw[->] (4) edge[loop below] (4);
\draw[->] (5) edge[loop below] (5);
\end{tikzpicture}
\end{center}
From the dependency pairs of $R$:
\begin{xalignat*}{4}
&(1)&
\m{primes}^{\#} &\to \m{sieve}^{\#}(\m{from}(\m{s}(\m{s}(\m{0})))) &
&(5)&
\m{sieve}^{\#}(\m{0}:y) &\to \m{sieve}^{\#}(y) \\
&(2)&
\m{primes}^{\#}\ &\to \m{from}^{\#}(\m{s}(\m{s}(\m{0}))) &
&(6)&
\m{sieve}^{\#}(\m{s}(n):y)\ &\to \m{sieve}^{\#}(\m{filter}(n,y,n)) \\
&(3)&
\m{from}^{\#}(n) &\to \m{from}^{\#}(\m{s}(n)) &
&(7)&
\m{sieve}^{\#}(\m{s}(n):y)\ &\to \m{filter}^{\#}(n,y,n) \\
&(4)&
\m{take}^{\#}(\m{s}(n),x:y) &\to \m{take}^{\#}(n,y) &
&(8)&
\m{filter}^{\#}(\m{0},x:y,m) &\to \m{filter}^{\#}(m,y,m) \\
&&
&&
&(9)&
\m{filter}^{\#}(\m{s}(n),x:y,m) &\to \m{filter}^{\#}(n,y,m)
\end{xalignat*}
the dependency graph can be constructed, where there is an edge from
$\ell_1 \to r_1$ to $\ell_2 \to r_2$
if $r_1$ may rewrite to a term matching $\ell_2$.
It provides a more fine-grained analysis than the previous graph, just like 
a basic block flow graph shows more details than a call graph.
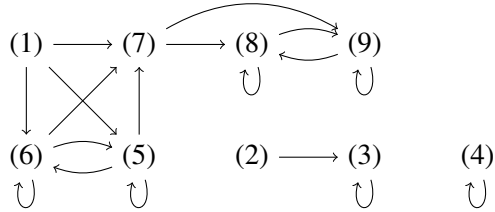
\begin{center}
\begin{tikzpicture}[node distance=15mm]
\node(1) {(1)};
\node[below of=1] (6) {(6)};
\node[right of=6] (5) {(5)};
\node[right of=5] (2) {(2)};
\node[right of=2] (3) {(3)};
\node[right of=3] (4) {(4)};
\node[right of=1] (7) {(7)};
\node[right of=7] (8) {(8)};
\node[right of=8] (9) {(9)};
\draw[->] (1) -- (5);
\draw[->] (1) -- (6);
\draw[->] (1) -- (7);
\draw[->] (2) -- (3);
\draw[->] (3) edge[loop below] (3);
\draw[->] (4) edge[loop below] (4);
\draw[->] (5) -- (7);
\draw[->] (5) edge[loop below] (5);
\draw[->] (5) edge[bend left=20] (6);
\draw[->] (6) edge[bend left=20] (5);
\draw[->] (6) edge[loop below] (6);
\draw[->] (6) -- (7);
\draw[->] (7) -- (8);
\draw[->] (7) edge[bend left]  (9);
\draw[->] (8) edge[loop below] (8);
\draw[->] (9) edge[bend left=20] (8);
\draw[->] (8) edge[bend left=20] (9);
\draw[->] (9) edge[loop below] (9);
\end{tikzpicture}
\end{center}
\end{example}

\smallskip
Finally, we mention some further features of term rewrite systems which
express structural properties that may be relevant. 
These include orthogonality, the number of critical pairs, or whether it is
a constructor system. But also more complex properties such as simple (non-)termination and local
(non-)confluence can be easily decided for some cases, and might be
indicative for the complexity of an input TRS and hence appropriate strategies.
Moreover, also the presence of algebraic substructures such as associative and
commutative operators, groups, lattices, or distributivity can be taken into account.

\section{Metrics for Theorem Proving Problems}
\label{sec:tp}

Here we consider theorem proving problems as collected in the TPTP benchmarks~\cite{TPTP}.
This collection is quite diverse. Given a theorem proving problem, it thus
seems natural to consider membership of this problem in different syntactic
problem classes, as in the case of rewrite systems.
In the case of TPTP problems, one might consider the unit equality class (UEQ),
the Bernays-Sch\"onfinkel class (EPR)~\cite{Le80}, higher-order classes, or
decidable first-order fragments like the monadic, the Ackermann, or the 
guarded fragment~\cite{ANB98},
description logics, or problems where symbols can be sorted in an acyclic
way~\cite{Ko13}.

Membership of the problem in classes like UEQ or EPR is already indicated
in the metadata of TPTP files, as are in fact many other syntactic properties:
the number
of (unit) clauses, literals, and terms; the number of occurrences of different
logical operators; the number of function symbols, predicates, and
(existentially and universally quantified) variables; formula and term depth;
as well as information about arithmetic operators and types. In learning
suitable strategies for a given problem, it cannot hurt to include these
features, since non-influential problem properties can be ignored by
the model.

Next, we mention some {\semantic} properties which can be included
to classify problems.
In presence of equality literals, all features described in
Section~\ref{sec:trs} can be considered for theorem proving problems as well.
Besides, the presence of substructures like group, list, lattice, or array axioms can be exploited.

For typed theorem proving problems, we can moreover take the number of
arguments of boolean, integer, real, and set type in every function symbol and
predicate into account. These features would resemble variable types being taken
into account as done in~\cite{DPVZ17}.

In \secref{trs} the dependency graph was used to keep track of dependencies
between function symbols.
A similar approach could be used to analyze the dependencies between
literals. For problems in CNF without equality, the 
following notion of a \depgraph could be used for this purpose.

\begin{definition}
Let $S$ be a set of variable-disjoint clauses without equality.
The \emph{\depgraph} has as node set all pairs of literals
$(l,l')$ such that there is a clause in $S$ containing $\neg l$ and
$l'$, and an edge from $(l_1,l_1')$ to $(l_2,l_2')$ if and only if $l_1'$
and $l_2$ have the same polarity and are unifiable
(where we identify literals $\neg \neg p$ and $p$, for all atoms $p$.)
\end{definition}

The \depgraph captures opportunities for resolution.
For instance, if this graph has no cycles then the problem admits only
finitely many resolution steps and is hence decidable.
We illustrate this idea by means of an example.
\begin{example}
Consider the following two sets of clauses
\begin{xalignat*}{4}
S_1\colon\quad&&
&\BLUE{\neg \m{P}(x) \vee \m{Q}(x)} &
&\GREEN{\neg \m{Q}(y) \vee \neg \m{R}(\m{f}(y), y)} &
&\YELLOW{\m{R}(u,z) \vee \m{P}(\m{f}(u))} 
\\
S_2\colon\quad&&
&\BLUE{\neg \m{P}(x) \vee \m{Q}(x)} &
&\GREEN{\neg \m{Q}(y) \vee \neg \m{R}(\m{f}(y), y)} &
&\RED{\m{R}(u,u) \vee \m{P}(\m{f}(u))}
\end{xalignat*}
The \depgraphs look as follows, where we use
colors to indicate from which clauses the literal pairs originate:
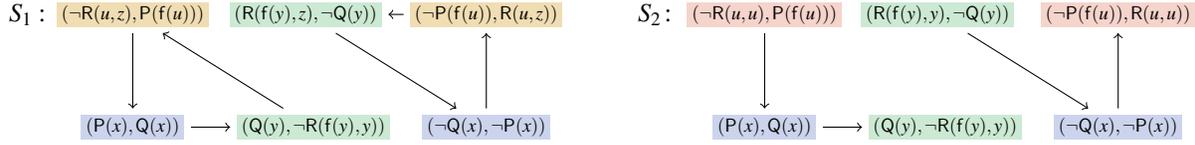
\begin{center}
\begin{tikzpicture}[node distance=14mm]
\tikzstyle{lits}=[scale=.65]
\tikzstyle{arr}=[->]
\pgfmathsetmacro{\cubex}{2.4}
\pgfmathsetmacro{\cubey}{1.5}
\pgfmathsetmacro{\shift}{2.3}
\node[lits] (10) at (0,0) {$\BLUE{(\m{P}(x),\m{Q}(x))}$};
\node[lits] (20) at (\cubex,0) {$\GREEN{(\m{Q}(y), \neg \m{R}(\m{f}(y),y))}$};
\node[lits] (30) at (0, \cubey) {$\YELLOW{(\neg \m{R}(u, z), \m{P}(\m{f}(u)))}$}
   node[left of=30, scale=.9] {$S_1:$};
\node[lits] (11) at (\cubex + \shift, 0) {$\BLUE{(\neg \m{Q}(x), \neg\m{P}(x))}$};
\node[lits] (21) at (\shift,\cubey) {$\GREEN{(\m{R}(\m{f}(y),z), \neg \m{Q}(y))}$};
\node[lits] (31) at (\cubex + \shift,\cubey) {$\YELLOW{(\neg \m{P}(\m{f}(u)), \m{R}(u, z))}$};
\draw[arr] (10) -- (20);
\draw[arr] (20) -- (30);
\draw[arr] (30) -- (10);
\draw[arr] (21) -- (11);
\draw[arr] (11) -- (31);
\draw[arr] (31) -- (21);
\begin{scope}[xshift=8.4cm]
\node[lits] (10) at (0,0) {$\BLUE{(\m{P}(x), \m{Q}(x))}$};
\node[lits] (20) at (\cubex,0) {$\GREEN{(\m{Q}(y), \neg \m{R}(\m{f}(y),y))}$};
\node[lits] (30) at (0, \cubey) {$\RED{(\neg \m{R}(u, u), \m{P}(\m{f}(u)))}$}
   node[left of=30, scale=.9] {$S_2\colon$};
\node[lits] (11) at (\cubex + \shift, 0) {$\BLUE{(\neg \m{Q}(x), \neg\m{P}(x))}$};
\node[lits] (21) at (\shift,\cubey) {$\GREEN{(\m{R}(\m{f}(y),y), \neg \m{Q}(y))}$};
\node[lits] (31) at (\cubex + \shift,\cubey) {$\RED{(\neg \m{P}(\m{f}(u)), \m{R}(u, u))}$};
\draw[arr] (10) -- (20);
\draw[arr] (30) -- (10);
\draw[arr] (21) -- (11);
\draw[arr] (11) -- (31);
\end{scope}
\end{tikzpicture}
\end{center}
The graph for $S_1$ has cycles and the accumulated substitutions in the cycles are
not just renamings. Indeed, $S_1$ offers infinitely many opportunities for resolution.
The graph for $S_2$ on the other hand is free of cycles, hence its saturation
is finite.
\end{example}

Properties of the \depgraph such as its number of nodes, degrees of
nodes, and cycles can thus be used as features indicating the proof complexity.
One could also consider to include further properties about these cycles
related to the accumulated substitution.
\smallskip

All the features mentioned so far in Sections~\ref{sec:trs} and~\ref{sec:tp}
were described as \emph{static} properties, in the sense that they
are computed from the input problem before starting a proof attempt.
They could also be considered as \emph{dynamic} features, meaning
they are re-evaluated after a certain number of proof steps.
This idea was already used in~\cite{BHP14}.
However, besides revising properties already considered for the initial
problem, also characteristics of the proof steps themselves could be
considered.

\section{Conclusion}
\label{sec:conclusion}

In this extended abstract we proposed {\semantic} features that can be used in
machine learning tasks to infer appropriate strategies or tools for
term rewriting or theorem proving problems. 

We believe that the development of meaningful features is interesting in 
multiple respects.
One aim is to use machine learning to better predict suitable strategies and
tools for a given problem.
But once a machine learning model is obtained, it can also be
analyzed for its most significant features. 
If a nontrivial feature turns out to be important, it constitutes
also an interesting property of problems that helps to discriminate between
different problem classes.
This should allow for interesting conclusions about which characteristics are
pivotal in the choice of a strategy or a strategy component, and consequently
lead to a taxonomy of problems.

The properties described here are, however, not meant to be comprehensive.
On the contrary, they rather constitute a collection of initial ideas.
In future work, we aim to test the expressibility of the proposed feature 
set experimentally, also attempting to complement it with
further useful properties.

\bibliographystyle{eptcs}
\bibliography{references}

\begin{thebibliography}{10}
\providecommand{\bibitemdeclare}[2]{}
\providecommand{\surnamestart}{}
\providecommand{\surnameend}{}
\providecommand{\urlprefix}{Available at }
\providecommand{\url}[1]{\texttt{#1}}
\providecommand{\href}[2]{\texttt{#2}}
\providecommand{\urlalt}[2]{\href{#1}{#2}}
\providecommand{\doi}[1]{doi:\urlalt{http://dx.doi.org/#1}{#1}}
\providecommand{\bibinfo}[2]{#2}

\bibitemdeclare{article}{ANB98}
\bibitem{ANB98}
\bibinfo{author}{H.~\surnamestart Andr\'{e}ka\surnameend},
  \bibinfo{author}{I.~\surnamestart N\'{e}meti\surnameend} \&
  \bibinfo{author}{J.~\surnamestart van Benthem\surnameend}
  (\bibinfo{year}{1998}): \emph{\bibinfo{title}{Modal logics and bounded
  fragments of predicate logic}}.
\newblock {\sl \bibinfo{journal}{JPL}}
  \bibinfo{volume}{27}(\bibinfo{number}{3}), pp. \bibinfo{pages}{217--274},
  \doi{10.1023/A:1004275029985}.

\bibitemdeclare{article}{AG00}
\bibitem{AG00}
\bibinfo{author}{T.~\surnamestart Arts\surnameend} \&
  \bibinfo{author}{J.~\surnamestart Giesl\surnameend} (\bibinfo{year}{2000}):
  \emph{\bibinfo{title}{Termination of term rewriting using dependency pairs}}.
\newblock {\sl \bibinfo{journal}{Theoretical Computer Science}}
  \bibinfo{volume}{236}(\bibinfo{number}{1}), pp. \bibinfo{pages}{133--178},
  \doi{10.1016/S0304-3975(99)00207-8}.

\bibitemdeclare{article}{AEM15}
\bibitem{AEM15}
\bibinfo{author}{M.~\surnamestart Avanzini\surnameend},
  \bibinfo{author}{N.~\surnamestart Eguchi\surnameend} \&
  \bibinfo{author}{G.~\surnamestart Moser\surnameend} (\bibinfo{year}{2015}):
  \emph{\bibinfo{title}{A new order-theoretic characterisation of the polytime
  computable functions}}.
\newblock {\sl \bibinfo{journal}{Theoretical Computer Science}}
  \bibinfo{volume}{585}, pp. \bibinfo{pages}{3--24},
  \doi{10.1016/j.tcs.2015.03.003}.

\bibitemdeclare{article}{AM13}
\bibitem{AM13}
\bibinfo{author}{M.~\surnamestart Avanzini\surnameend} \&
  \bibinfo{author}{G.~\surnamestart Moser\surnameend} (\bibinfo{year}{2013}):
  \emph{\bibinfo{title}{Polynomial Path Orders}}.
\newblock {\sl \bibinfo{journal}{Log.\ Meth.\ Comput.\ Sci.}}
  \bibinfo{volume}{9}(\bibinfo{number}{4}), \doi{10.2168/LMCS-9(4:9)2013}.

\bibitemdeclare{book}{BN98}
\bibitem{BN98}
\bibinfo{author}{F.~\surnamestart Baader\surnameend} \&
  \bibinfo{author}{T.~\surnamestart Nipkow\surnameend} (\bibinfo{year}{1998}):
  \emph{\bibinfo{title}{Term Rewriting and All That}}.
\newblock \bibinfo{publisher}{Cambridge University Press},
  \doi{10.1017/CBO9781139172752}.

\bibitemdeclare{article}{BellantoniCook:1992}
\bibitem{BellantoniCook:1992}
\bibinfo{author}{S.~\surnamestart Bellantoni\surnameend} \&
  \bibinfo{author}{S.~\surnamestart Cook\surnameend} (\bibinfo{year}{1992}):
  \emph{\bibinfo{title}{A new recursion-theoretic characterization of the
  polytime functions}}.
\newblock {\sl \bibinfo{journal}{Comput. Complex.}}
  \bibinfo{volume}{2}(\bibinfo{number}{2}), pp. \bibinfo{pages}{97--110},
  \doi{10.1145/129712.129740}.

\bibitemdeclare{article}{Ben-Amram:2011}
\bibitem{Ben-Amram:2011}
\bibinfo{author}{A.~\surnamestart Ben-Amram\surnameend} (\bibinfo{year}{2011}):
  \emph{\bibinfo{title}{Monotonicity Constraints for Termination in the Integer
  Domain}}.
\newblock {\sl \bibinfo{journal}{Log.\ Meth.\ Comput.\ Sci.}}
  \bibinfo{volume}{7}(\bibinfo{number}{3}), \doi{10.2168/LMCS-7(3:4)2011}.

\bibitemdeclare{inproceedings}{BP:2016}
\bibitem{BP:2016}
\bibinfo{author}{A.~\surnamestart Ben{-}Amram\surnameend} \&
  \bibinfo{author}{A.~\surnamestart Pineles\surnameend} (\bibinfo{year}{2016}):
  \emph{\bibinfo{title}{Flowchart Programs, Regular Expressions, and
  Decidability of Polynomial Growth-Rate}}.
\newblock In: {\sl \bibinfo{booktitle}{Proc.\ 4th VPT}}, {\sl
  \bibinfo{series}{EPTCS}} \bibinfo{volume}{216}, pp. \bibinfo{pages}{24--49},
  \doi{10.4204/EPTCS.216.2}.

\bibitemdeclare{article}{BHP14}
\bibitem{BHP14}
\bibinfo{author}{J.~P. \surnamestart Bridge\surnameend},
  \bibinfo{author}{S.~\surnamestart Holden\surnameend} \&
  \bibinfo{author}{L.~\surnamestart Paulson\surnameend} (\bibinfo{year}{2014}):
  \emph{\bibinfo{title}{Machine Learning for First-Order Theorem Proving}}.
\newblock {\sl \bibinfo{journal}{JAR}}
  \bibinfo{volume}{53}(\bibinfo{number}{2}), pp. \bibinfo{pages}{141--172},
  \doi{10.1007/s10817-014-9301-5}.

\bibitemdeclare{inproceedings}{CJSU19}
\bibitem{CJSU19}
\bibinfo{author}{K.~\surnamestart Chvalovsk\'y\surnameend},
  \bibinfo{author}{J.~\surnamestart Jakub\r{u}v\surnameend},
  \bibinfo{author}{M.~\surnamestart Suda\surnameend} \&
  \bibinfo{author}{J.~\surnamestart Urban\surnameend} (\bibinfo{year}{2019}):
  \emph{\bibinfo{title}{{ENIGMA-NG:} Efficient Neural and Gradient-Boosted
  Inference Guidance for {E}}}.
\newblock In: {\sl \bibinfo{booktitle}{Proc.\ 27th CADE}}, {\sl
  \bibinfo{series}{LNCS}} \bibinfo{volume}{11716}, pp.
  \bibinfo{pages}{197--215}, \doi{10.1007/978-3-030-29436-6\_12}.

\bibitemdeclare{inproceedings}{CDZ:2017}
\bibitem{CDZ:2017}
\bibinfo{author}{T.~\surnamestart Colcombet\surnameend},
  \bibinfo{author}{L.~\surnamestart Daviaud\surnameend} \&
  \bibinfo{author}{F.~\surnamestart Zuleger\surnameend} (\bibinfo{year}{2017}):
  \emph{\bibinfo{title}{Automata and Program Analysis}}.
\newblock In: {\sl \bibinfo{booktitle}{Proc.\ 21st FCT}}, {\sl
  \bibinfo{series}{LNCS}} \bibinfo{volume}{10472}, pp. \bibinfo{pages}{3--10},
  \doi{10.1007/978-3-662-55751-8\_1}.

\bibitemdeclare{inproceedings}{DT90}
\bibitem{DT90}
\bibinfo{author}{M.~\surnamestart Dauchet\surnameend} \&
  \bibinfo{author}{S.~\surnamestart Tison\surnameend} (\bibinfo{year}{1990}):
  \emph{\bibinfo{title}{The Theory of Ground Rewrite Systems is Decidable}}.
\newblock In: {\sl \bibinfo{booktitle}{Proc.\ 5thIEEE Symposium on Logic in
  Computer Science}}, pp. \bibinfo{pages}{242--248},
  \doi{10.1109/LICS.1990.113750}.

\bibitemdeclare{article}{DPVZ17}
\bibitem{DPVZ17}
\bibinfo{author}{Y.~\surnamestart Demyanova\surnameend},
  \bibinfo{author}{T.~\surnamestart Pani\surnameend},
  \bibinfo{author}{H.~\surnamestart Veith\surnameend} \&
  \bibinfo{author}{F.~\surnamestart Zuleger\surnameend} (\bibinfo{year}{2017}):
  \emph{\bibinfo{title}{Empirical Software Metrics for Benchmarking of
  Verification Tools}}.
\newblock {\sl \bibinfo{journal}{Form. Methods Syst. Des.}}
  \bibinfo{volume}{50}(\bibinfo{number}{2-3}), pp. \bibinfo{pages}{289--316},
  \doi{10.1007/s10703-016-0264-5}.

\bibitemdeclare{inproceedings}{JU17}
\bibitem{JU17}
\bibinfo{author}{J.~\surnamestart Jakub\r{u}v\surnameend} \&
  \bibinfo{author}{J.~\surnamestart Urban\surnameend} (\bibinfo{year}{2017}):
  \emph{\bibinfo{title}{{ENIGMA:} Efficient Learning-Based Inference Guiding
  Machine}}.
\newblock In: {\sl \bibinfo{booktitle}{Proc.\ CICM 2017}}, {\sl
  \bibinfo{series}{LNCS}} \bibinfo{volume}{10383}, pp.
  \bibinfo{pages}{292--302}, \doi{10.1007/978-3-319-62075-6\_20}.

\bibitemdeclare{article}{JU18}
\bibitem{JU18}
\bibinfo{author}{J.~\surnamestart Jakub\r{u}v\surnameend} \&
  \bibinfo{author}{J.~\surnamestart Urban\surnameend} (\bibinfo{year}{2018}):
  \emph{\bibinfo{title}{Hierarchical invention of theorem proving strategies}}.
\newblock {\sl \bibinfo{journal}{{AI} Commun.}}
  \bibinfo{volume}{31}(\bibinfo{number}{3}), pp. \bibinfo{pages}{237--250},
  \doi{10.3233/AIC-180761}.

\bibitemdeclare{inproceedings}{KU15}
\bibitem{KU15}
\bibinfo{author}{C.~\surnamestart Kaliszyk\surnameend} \&
  \bibinfo{author}{J.~\surnamestart Urban\surnameend} (\bibinfo{year}{2015}):
  \emph{\bibinfo{title}{{FEMaLeCoP}: Fairly Efficient Machine Learning
  Connection Prover}}.
\newblock In: {\sl \bibinfo{booktitle}{Proc.\ LPAR 2015}}, {\sl
  \bibinfo{series}{LNCS}} \bibinfo{volume}{9450}, pp. \bibinfo{pages}{88--96},
  \doi{10.1007/978-3-662-48899-7\_7}.

\bibitemdeclare{inproceedings}{Ko13}
\bibitem{Ko13}
\bibinfo{author}{K.~\surnamestart Korovin\surnameend} (\bibinfo{year}{2013}):
  \emph{\bibinfo{title}{Non-cyclic Sorts for First-Order Satisfiability}}.
\newblock In: {\sl \bibinfo{booktitle}{Proc.\ FroCoS 2013}}, {\sl
  \bibinfo{series}{LNCS}} \bibinfo{volume}{8152}, pp.
  \bibinfo{pages}{214--228}, \doi{10.1007/978-3-642-40885-4\_15}.

\bibitemdeclare{inproceedings}{KSU13}
\bibitem{KSU13}
\bibinfo{author}{D.~\surnamestart K{\"u}hlwein\surnameend},
  \bibinfo{author}{S.~\surnamestart Schulz\surnameend} \&
  \bibinfo{author}{J.~\surnamestart Urban\surnameend} (\bibinfo{year}{2013}):
  \emph{\bibinfo{title}{{E-MaLeS} 1.1}}.
\newblock In: {\sl \bibinfo{booktitle}{Proc.\ 24th CADE}}, {\sl
  \bibinfo{series}{LNCS}} \bibinfo{volume}{7898}, pp.
  \bibinfo{pages}{407--413}, \doi{10.1007/978-3-642-38574-2}.

\bibitemdeclare{incollection}{Leivant:1994}
\bibitem{Leivant:1994}
\bibinfo{author}{D.~\surnamestart Leivant\surnameend} (\bibinfo{year}{1994}):
  \emph{\bibinfo{title}{Ramified recurrence and computatinal complexity {I}:
  {W}ord recurrence and poly-time}}.
\newblock In \bibinfo{editor}{P.~\surnamestart Clote\surnameend} \&
  \bibinfo{editor}{J.~\surnamestart Remmel\surnameend}, editors: {\sl
  \bibinfo{booktitle}{Feasible Mathematics {II}}},
  \bibinfo{publisher}{Birkh{\"a}user}, pp. \bibinfo{pages}{320--343},
  \doi{10.1007/978-1-4612-2566-9\_11}.

\bibitemdeclare{article}{Le80}
\bibitem{Le80}
\bibinfo{author}{H.~R. \surnamestart Lewis\surnameend} (\bibinfo{year}{1980}):
  \emph{\bibinfo{title}{Complexity results for classes of quantificational
  formulas}}.
\newblock {\sl \bibinfo{journal}{JCSS}}
  \bibinfo{volume}{21}(\bibinfo{number}{3}), pp. \bibinfo{pages}{317--353},
  \doi{10.1016/0022-0000(80)90027-6}.

\bibitemdeclare{inproceedings}{LISK17}
\bibitem{LISK17}
\bibinfo{author}{S.~M. \surnamestart Loos\surnameend},
  \bibinfo{author}{G.~\surnamestart Irving\surnameend},
  \bibinfo{author}{C.~\surnamestart Szegedy\surnameend} \&
  \bibinfo{author}{C.~\surnamestart Kaliszyk\surnameend}
  (\bibinfo{year}{2017}): \emph{\bibinfo{title}{Deep Network Guided Proof
  Search}}.
\newblock In: {\sl \bibinfo{booktitle}{Proc. 21st LPAR}}, pp.
  \bibinfo{pages}{85--105}, \doi{10.29007/8mwc}.

\bibitemdeclare{inproceedings}{Sch01}
\bibitem{Sch01}
\bibinfo{author}{S.~\surnamestart Schulz\surnameend} (\bibinfo{year}{2001}):
  \emph{\bibinfo{title}{Learning Search Control Knowledge for Equational
  Theorem Proving}}.
\newblock In: {\sl \bibinfo{booktitle}{Proc.\ KI 2001}}, {\sl
  \bibinfo{series}{LNCS}} \bibinfo{volume}{2174}, pp.
  \bibinfo{pages}{320--334}, \doi{10.1007/3-540-45422-5\_23}.

\bibitemdeclare{article}{Si88}
\bibitem{Si88}
\bibinfo{author}{H.~\surnamestart Simmons\surnameend} (\bibinfo{year}{1988}):
  \emph{\bibinfo{title}{The realm of primitive recursion}}.
\newblock {\sl \bibinfo{journal}{Archive for Mathematical Logic}}
  \bibinfo{volume}{27}(\bibinfo{number}{2}), pp. \bibinfo{pages}{177--188},
  \doi{10.1007/BF01620765}.

\bibitemdeclare{article}{TPTP}
\bibitem{TPTP}
\bibinfo{author}{G.~\surnamestart Sutcliffe\surnameend} (\bibinfo{year}{2009}):
  \emph{\bibinfo{title}{{The TPTP Problem Library and Associated
  Infrastructure: The FOF and CNF Parts}}}.
\newblock {\sl \bibinfo{journal}{JAR}}
  \bibinfo{volume}{43}(\bibinfo{number}{4}), pp. \bibinfo{pages}{337--362},
  \doi{10.1007/s10817-009-9143-8}.

\end{thebibliography}

\end{document}